\documentclass[11pt]{article}
\usepackage{coling2020}
\usepackage{times}
\usepackage{url}
\usepackage{latexsym}

\usepackage{array}
\usepackage[ruled,vlined]{algorithm2e}
\usepackage{amsfonts}
\usepackage{amsmath}
\usepackage{amssymb}
\usepackage{amsthm}
\usepackage{bm}
\usepackage{booktabs}
\usepackage{color}
\usepackage{graphicx}
\usepackage{multirow}
\usepackage{subfigure}
\usepackage{wrapfig}

\usepackage{accents}

\colingfinalcopy 

\title{Embedding Dynamic Attributed Networks by Modeling the Evolution Processes}

\author{
	Zenan Xu$^{1,2}$,~ Zijing Ou$^1$,~ Qinliang  Su$^{1,2,3}$\thanks{~~Corresponding author.},~ Jianxing Yu$^1$, \\\bf Xiaojun Quan$^1$ and Zhenkun Lin$^{1,2}$\\
	$^1$School of Data and Computer Science, Sun Yat-sen University \\
	$^2$Guangdong Key Laboratory of Big Data Analysis and Processing, Guangzhou, China\\
	$^3$Key Lab. of Machine Intelligence and Advanced	Computing, Ministry of Education, China
	\\
	\texttt{\{xuzn,~ouzj,~linzhk3\}@mail2.sysu.edu.cn}\\
	\texttt{\{suqliang,~yujx,~quanxj3\}@mail.sysu.edu.cn}
}

\date{}

\begin{document}
\maketitle
\begin{abstract}
  Network embedding has recently emerged as a promising technique to embed nodes of a network into low-dimensional vectors. While fairly successful, most existing works focus on the embedding techniques for static networks. But in practice, there are many networks that are evolving over time and hence are dynamic, {\it e.g.}, the social networks. To address this issue, a high-order spatio-temporal embedding model is developed to track the evolutions of dynamic networks. Specifically, an activeness-aware neighborhood embedding method is first proposed to extract the high-order neighborhood information at each given timestamp. Then, an embedding prediction framework is further developed to capture the temporal correlations, in which the attention mechanism is employed instead of recurrent neural networks (RNNs) for its efficiency in computing and flexibility in  modeling. Extensive experiments are conducted on four real-world datasets from three different areas. It is shown that the proposed method outperforms all the baselines by a substantial margin for the tasks of dynamic link prediction and node classification, which demonstrates the effectiveness of the proposed methods on tracking the evolutions of dynamic networks.
\end{abstract}

\section{Introduction}

Network embedding (NE) aims to represent each node by a low-dimensional vector, while seeking to preserve their neighborhood information as much as possible. It has been shown that working on the low-dimensional representations is much more efficient than on the original large-scale networks directly in various real-world applications, such as  friend recommendation, product advertising, community detection \cite{cavallari2017learning},  nodes classification {\it etc}. Because of its capability in facilitating downstream applications, many methods have been developed to embed network nodes into vectors efficiently and effectively, like DeepWalk in \cite{perozzi2014deepwalk}, LINE in \cite{tang2015line}, Node2Vec in \cite{grover2016node2vec} {\it etc}. Later, the attributes/texts available at nodes are further taken into account, {\it e.g.} the CANE in \cite{tu2017cane} and WANE in \cite{shen2018improved}, to obtain more comprehensive embeddings. However, these methods mostly focus on static networks, but in practice, networks are often dynamic. In social networks, for instance, new friend connections are established all the time, and user profiles are also updated from time to time. For these dynamic networks, how to learn their embeddings, and more importantly, how to leverage the embeddings to predict their evolution trends is crucial for many applications.

Existing methods for dynamic network embedding can be roughly divided into two categories. The first category concerns about how to obtain new embeddings from the stale ones efficiently when changes of networks are observed. In \cite{du2018dynamic}, by decomposing the learning objective into different parts, it is shown that new embeddings can be updated by only considering the newly added and most influential nodes. Differently, \cite{hamilton2017inductive,cheng2020dynamic} proposed to use a graph convolutional network (GCN) or Gaussian process to learn a mapping from the associated attributes to the embeddings, respectively, then the embeddings can be updated directly with the output 
\begin{wrapfigure}{r}{8cm}
	\centering
	\includegraphics[scale=1.15]{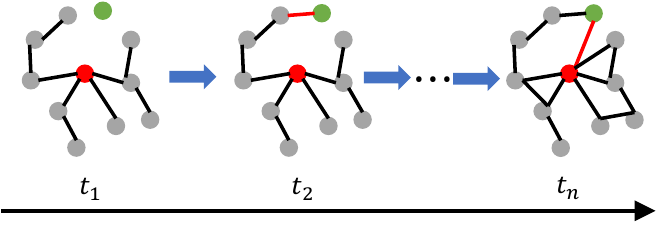}
	\caption{Illustration of the evolution process of a dynamic network, where red lines indicate the newly formed relationships.}
	\label{fig:intro}
\end{wrapfigure}
of the mapping. The aforementioned methods avoid to re-compute the embeddings from scratch at every timestamp by tracking the network changes continuously. However, since these methods focus only on the changes at the current timestamp rather than their dynamics, the induced embeddings can only represent the networks at the current timestamp, but are poor at predicting their future evolvement.


The second category methods focus more on the improvement of prediction performance. \newcite{singer2019node} proposed to learn embeddings and an alignment matrix for the network at each timestamp by solving a sequential optimization problem, and then input the  embeddings into recurrent neural networks (RNN) to predict the future links. However, sequential optimization is very expensive, hindering it from being applied to large networks. Alternatively, \newcite{zhou2018dynamic}  proposed to learn embeddings with the objective to predict the future closure process of nodes that are separated by at most two hops. Later, \newcite{goyal2020dyngraph2vec} proposed to employ auto-encoders to predict nodes' direct neighbors with the neighbors from the previous timestamps via the long short-term memory (LSTM). Recently, \newcite{zuo2018embedding} introduced the concept of neighborhood formation and used it to track the evolution of nodes with their direct neighbors from the previous timestamps. In all of these methods, only the direct (first-order) neighbors in the spatial dimension are leveraged. However, to capture the dynamics of networks, it is  important to consider the high-order information of nodes in both the spatial and temporal dimensions simultaneously. As illustrated in Fig.\ref{fig:intro}, the green node is isolated from the red node at $t_1$ and becomes its fourth-order neighbor at $t_2$, and further evolves into its direct neighbor at $t_n$. To capture this evolution process, the model should have the ability to be aware of nodes' high-order neighborhood information spatially and memorize the changes that occurred many timestamps before temporally.

In this paper, a dynamic attribute network embedding model (Dane) is developed to track the evolutions of dynamic networks. Specifically, an activeness-aware neighborhood embedding method is proposed to extract the high-order neighborhood information at each given timestamp. The activeness-aware mechanism enables the model to emphasize more on nodes that are active in social activities. Then, an embedding prediction framework is developed to capture the temporal correlations of dynamic networks, in which the attention mechanism is employed instead of RNNs for its efficiency in computing and flexibilities in modeling. The methods are evaluated on the tasks of dynamic link prediction and node classification over four real-world datasets. It is shown that the proposed methods outperform all comparable embedding methods, including both static and dynamic ones, on the link prediction by a substantial margin. This demonstrates that the proposed methods are able to capture the correlations from the dynamic networks. Similar phenomena are observed on the task of dynamic node classification, which further confirms the effectiveness of the proposed methods on tracking the evolutions of dynamic networks. 

\section{Related Work}
Representation learning for graphs has attracted considerable attention recently, since they can potentially benefit a wide range of applications. Specifically, \cite{perozzi2014deepwalk} employs random walks to obtain sequences of nodes, and then uses the word2vec technique to represent the nodes into low-dimensional vectors.  To preserve both the global and local structure information, \cite{tang2015line} proposed to jointly optimize the first- and second-order proximity of nodes in network. Later, \cite{grover2016node2vec} introduced a biased random walk procedure under the BFS and DFS search strategies to further exploit the diversity of structure patterns of networks. However, in all of these methods, only the topology information of network is leveraged. But in many real-world networks, nodes are often associated with attributes or texts. To take the attributes into account, a mutual attention \begin{wrapfigure}{r}{8cm}
	\centering
	\includegraphics[scale=0.95]{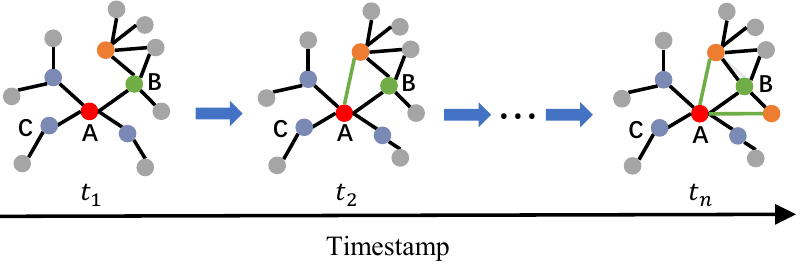}
	\caption{Illustration of how the activeness of an node effects the evolution of network.}
	\label{fig:3.1}
\end{wrapfigure}
mechanism is designed to enhance the relationships between attributes of neighboring nodes in \cite{tu2017cane}. To extract more semantic features from attributed data, \cite{shen2018improved,xu2019deep} modified the mutual attention mechanism in \cite{tu2017cane} and employed a fine-grained word alignment mechanism.  However, all these works are overwhelmingly performed in the context of plain or static networks.

Various techniques have been proposed to learn deep representations in dynamic patterns. \cite{li2017attributed} first provides an offline embedding method and then leverages matrix perturbation theory to maintain the freshness of the end embedding results in an online manner. In \cite{seo2018structured}, the model was developed to combine convolutional neural networks on graphs to identify spatial structures and RNN to find dynamic patterns. Similarly, \cite{trivedi2017know} proposed a deep recurrent architecture to model historical evolution of entity embeddings in a specific relationship space. To capture both structural properties and temporal evolutionary patterns, \cite{sankar2018dynamic} jointly employed self-attention layers along structural neighborhood and temporal dynamics.  Differently, \cite{nguyen2018continuous} employed a temporal version of traditional random walks to capture temporally evolving neighborhood information. Although significant efforts have been made to learn effective representations in evolutionary patterns, methods aiming to extract high-order spatio-temporal evolutionary information have been rarely explored. In our work, an activeness-aware neighborhood embedding method is developed to capture high-order neighbor evolutionary relationship, followed by a low-complexity attention-based mechanism, which efficiently apprehends temporal evolutionary patterns.

\section{The Proposed Framework}
A dynamic attributed network $\mathcal{G}$ consists of a sequence of snapshots $\mathcal{G} \triangleq \{G_1,G_2,\dots,G_n\}$, where $G_t \triangleq (\bm{V}_t, \bm{E}_t, \bm{A}_t)$ is the attributed network at timestamp $t$; $\bm{V}_t$ and $\bm{E}_t$ are the set of vertices and edges in the network $G_t$, and $\bm{A}_t$ denotes the attribute matrix, with the $v$-th row representing the attribute associated with the node $v$. Additionally, we use $\bm V=\bm V_1 \cup ... \cup \bm V_n $ to denote the nodes of the whole network. In this section, we first propose a method to effectively extract the high-order neighborhood information at a given timestamp, based on which a low-complexity attention-based model is then developed to capture the network's future evolvement.

\subsection{Activeness-aware Neighborhood Embedding}\label{sec:SG}
To extract high-order neighborhood information in dynamic attributed networks, we can simply apply the GraphSAGE algorithm to the networks at each timestamp,  which  aggregates messages from neighboring nodes iteratively \cite{hamilton2017inductive}. However, all messages in GraphSAGE are treated equally. There is no problem if the embeddings are only used to represent the network at the current timestamp. But if they are used to predict future evolvement, it would be problematic. To see this, let us take the social network as an example. As illustrated in Fig \ref{fig:3.1}, suppose that both node B and C are the direct neighbors of node A, but node B is much more active than node C in organizing various social activities. Obviously, node A is more likely to establish connections with the friends of node B than node C in future. If the goal is to predict the future evolving trends, it is useful to know the activeness of different nodes, and more emphases should be given to the messages from the nodes that are more active.


To this end, an activeness-aware neighborhood embedding method is proposed. Specifically, given the attributed network ${G}_t=\{\bm{V}_t, \bm{E}_t, \bm{A}_t\}$ at timestamp $t$, the embedding of node $v$ is learned by updating the following equations
\begin{equation} \label{equ:S1}
\bar{\bm{x}}^{\ell}_{t, v} \!= \!{Mean}\left({Aggregate}(\bm{p}_{t,u}^{\ell}\odot \bm{x}^{\ell}_{t,u}, \forall u \in \mathcal{N}_t(v))\right),
\end{equation}
\begin{equation} \label{equ:S2}
\bm{x}^{\ell+1}_{t,v} = \tanh(\bm{W}_x^\ell\cdot[\bar{\bm{x}}^{\ell}_{t,v};\bm{x}^{\ell}_{t,v}]),
\end{equation}
where $\ell=0,1,\dots,L$; $\bm{x}^{\ell}_{t,u} \in {\mathbb{R}}^{ d}$ represents the embedding of node $u$ at the $\ell$-th layer , with $\bm{x}^{0}_{t,u}$ initialized with the $u$-th row of the attributed matrix ${\bm{A}}_t$; the values in $\bm{p}_{t,u}^{\ell}$ represent the degree of activeness of node $u$, which will be discussed in detail later; $\mathcal{N}_t(v)$ denotes the set of neighbors of node $v$ at the timestamp $t$;  the function ${Aggregate}(\cdot)$ collects messages $\bm{p}_{t,u}^{\ell}\odot \bm{x}^{\ell}_{t,u}$ from all neighbors $u \in \mathcal{N}_t(v)$ to constitute a matrix; ${Mean}(\cdot)$ means taking the average of a matrix along its rows; $\odot$ is element-wise multiplication; $[\cdot \; ; \; \cdot]$ means the concatenation of two vectors; and $\bm{W}_x^{\ell}$ is the parameters to be learned. From \eqref{equ:S1}, we can see that the activeness vector $\bm{p}_{t,u}^{\ell}$ play the role of gates. If the node $u$ is very active, the values of the corresponding activeness vector $\bm{p}_{t,u}^{\ell}$ will be large. Hence, a larger proportion of the embedding of node $u$ will be flowed into its neighboring nodes,  exerting a greater influence onto the other nodes. On the contrary, if the node $u$ is not active in social activities, its importance will be lowered by diminishing the values in $\bm{p}_{t,u}^{\ell}$.

For the activeness vectors $\bm{p}_{t,v}^{\ell}$, it is computed from a randomly initialized time-invariant matrix $\bm{P}\in\mathbb{R}^{|{\bm V}| \times d}$ as
\begin{equation}
\bar{\bm{p}}_{t,v}^\ell = Mean\left(Aggregate(\bm{p}_{t,u}^{\ell}, \forall u \in\mathcal{N}_t(v))\right),
\end{equation}
\begin{equation} \label{equ:S3}
\bm{p}^{\ell+1}_{t,v} = \sigma(\bm{W}_p^\ell\cdot[\bar{\bm{p}}^{\ell}_{t,v}; \bm{p}^{\ell}_{t,v}]),
\end{equation}
where $\sigma(\cdot)$ denotes the sigmoid function; and $\bm{p}_{t,v}^0$ is initialized by the $v$-th row of the matrix $\bm{P}$, which, together with the model parameter $\bm{W}_p^\ell$, is learned from the training data. Although $\bm{p}_{t,v}^{\ell}$ is computed from the time-invariant $\bm{P}$, because  its computation also depends on the time-evolving network topologies, the vector $\bm{p}_{t,v}^{\ell}$ can still track the evolvement of networks.


\subsection{Prediction of the Next-Timestamp Embedding}\label{sec:TG}
Given the embeddings ${\bm{x}}^\ell_{i, v}$ for $i=1, 2, \cdots, t$, in this section, we focus on how to predict the network status at the next timestamp $t+1$, $\it e.g.$, the link connections and node categories at $t+1$. In this paper, we predict the future network status by estimating the embeddings at the next-timestamp $\hat{\bm{x}}_{t+1, v}$ using the previous ones, that is, finding the mapping
\begin{equation}
{\mathcal{F}}: \{\bm{x}_{1,v}^\ell,\bm{x}_{2,v}^\ell,\cdots,\bm{x}_{t,v}^\ell\}_{\ell = 1}^L \rightarrow \hat{\bm{x}}_{t+1, v}.
\end{equation}
The most direct way is to feed the previous embeddings of each layer $\ell$ into an RNN or LSTM and then combine the predictions of different layers linearly as the final prediction, {\it i.e.},
\begin{align}
\hat{\bm{x}}_{t+1, v}^{\ell} &= RNN(\bm{x}_{1,v}^\ell,\bm{x}_{2,v}^\ell,\cdots,\bm{x}_{t,v}^\ell),	\\
\hat{\bm{x}}_{t+1, v} &= \frac{1}{L}\sum_{\ell=1}^{L}(\bm{W}_{y}\hat{\bm{x}}_{t+1,v}^\ell + \bm{b}_{y}), \label{pred_emb_merge}
\end{align}
where $\bm{W}_{y}$ and $\bm{b}_{y}$ are model parameters.  It can be seen from \eqref{equ:S1} and \eqref{equ:S2} that as the number of layer $\ell$ increases, broader neighborhood information will be included in the embeddings, but at the same time, the local information around each node will be weakened. Thus, to retain both the global and local neighborhood information, the embeddings $\bm{x}^{\ell}_{t,v}$ obtained from all intermediate layers $\ell = 1, 2, \cdots, L$ are employed for the embedding prediction.

RNNs or LSTMs are good at modeling the time dependencies of sequences, but their computations are also known to be time-consuming due to the difficulties of parallelizing. Actually, for many interesting dynamic networks, the changes are not dramatic for each timestamp. To better model the temporal correlation and speed up the computations, we further propose an attention-based model to predict the next-timestamp embedding as
\begin{equation} \label{equ:hw1}
\widetilde{\bm{x}}^\ell_{t,v}=\text{tanh}(\sum_{k=1}^{K}\alpha_{t-k,v}^\ell\bm{x}_{t-k, v}^\ell),
\end{equation}
\begin{equation} \label{equ:hw2}
\hat{\bm{x}}_{t+1,v}^\ell = \bm{x}_{t,v}^\ell +  \bm{g}_{t,v}^\ell \odot (\bm{x}_{t,v}^\ell - \widetilde{\bm{x}}_{t,v}^\ell),
\end{equation}
where $\widetilde{\bm{x}}^\ell_{t,v}$ denotes the summarized representation of the most recent $K$ embeddings until timestamp $t-1$; $K$ is the number of used historical embeddings; $\alpha_{t-k}^\ell$ is the attention coefficient and is computed as
\begin{equation} \label{equ:hw4}
\alpha_{t-k,v}^\ell=\frac{\text{exp}(\beta_{t-k,v}^\ell)}{\text{exp}(\sum_{k=1}^{K}\beta_{t-k,v}^\ell)},
\end{equation}
with the coefficient $
\beta_{t-k,v}^\ell =\sigma({(\bm{x}_{t-k,v}^\ell)}^T\bm{W}^\ell_\beta\bm{x}_{t-1,v}^\ell)$;
and the vector 
\begin{equation} \label{equ:hw3}
\bm{g}_{t,v}^\ell = \sigma(\bm{W}_{g}^\ell[\widetilde{\bm{x}}^\ell_{t,v};\bm{x}_{t,v}^\ell] + \bm{b}_{g}^\ell)
\end{equation}
controls how much of the change $\bm{x}_{t,v}^\ell - \widetilde{\bm{x}}_{t,v}^\ell$ at the previous timestamp are used for the prediction of the next timestamp.

\subsection{Training Objective}
Suppose there exists a link $e_{v,u}$ such that $e_{v,u}\in \bm{E}_{t+1}$ but $e_{v,u}\notin \{\bm{E}_1 \cup \bm{E}_2 \cup,\cdots,\cup \bm{E}_{t}\}$. If our proposed method is able to capture the historical dynamics and make good prediction for timestamp $t+1$, the prediction embedding for node $u$ and $v$, {\it i.e.}, $\hat{\bm{x}}_{t+1,v}$ and $\hat{\bm{x}}_{t+1,u}$, should be close to each other in the vector space. Thus, we define the objective function as as
\begin{equation} \label{l1}
\mathcal{L} = - \sum_{e_{v, u}\in {\bm E}_{t+1}} \log p(\hat{\bm{x}}_{t+1,v}|\hat{\bm{x}}_{t+1,u}),
\end{equation}
where $p(\hat{\bm{x}}_{t,v}|\hat{\bm{x}}_{t,u})$ denotes the conditional probability of embedding $\hat{\bm{x}}_{t,u}$ given the embedding $\hat{\bm{x}}_{t,v}$ and is defined as
\begin{equation}
p(\hat{\bm{x}}_{t,v}|\hat{\bm{x}}_{t,u}) \triangleq \frac{\text{exp}(\hat{\bm{x}}_{t,u}^T \hat{\bm{x}}_{t,v})}{\sum_{\{z,v\}\in {\bm E}_{t+1}} \text{exp}(\hat{\bm{x}}_{t,z}^T \hat{\bm{x}}_{t,v}))}.
\end{equation}
To alleviate the computational burden of repeatedly evaluating the softmax function, as done in \cite{mikolov2013distributed}, the negative sample technique is employed by optimizing the alternative loss below
\begin{align}
{\mathcal{\tilde L}} = - \log \sigma(\hat{\bm{x}}_{t+1,u}^T\hat{\bm{x}}_{t+1,v}) - \sum_{r=1}^{R}E_{z\sim D(z)}[\log \sigma (-\hat{\bm{x}}_{t+1,z}^T \hat{\bm{x}}_{t+1,v})],
\end{align}
where $R$ is the number of negative samples and $D(v) \propto d_v^{3/4}$ is the distribution of vertices, where $d_v$ denotes the out-degree of vertex $v$.

\section{Experiments}
In this section, we evaluate the performance of the proposed methods on two tasks: dynamic link prediction and node classification. For the link prediction, we predict the new links that appear at timestamp $t+1$ for the first time based on the historical observations until $t$ \cite{goyal2018dyngem}. For the task of node classification, the categories of nodes at timestamp $t+1$ are predicted, with only the nodes that change their categories at $t+1$ considered \cite{zhou2018dynamic}.  In the experiments, the dimension of network embeddings is set to 100 for all considered methods. The negative samples are set to 1 and the mini-batch size is set to 50 to speed up the training process. Adam\cite{kingma2014adam} is employed to train the proposed model with a learning rate of $1 \times 10^{-4}$.

\subsection{Datasets, Baselines and Evaluation Metric}
\paragraph{Datasets} To evaluate our proposed methods, we collect four real-world dynamic attributed network datasets, ranging from user action network, brain activity network to academic citation network. The statistics of four datasets are summarized in Table \ref{table:datasets}.

\begin{itemize}
	\item {\emph{MOOC}} is a user action dataset collected by \cite{kumar2019predicting}, in which users and course activities are represented as nodes, and actions by users on the course are represented as edges. The actions have attributes and timestamp, hence it can be recognized as a dynamic attributed graph. In our experiment, we split the dataset into 20 timestamps.
	
	\item {\emph{Brain}} is a brain activity dataset collected by \cite{xu2019spatio}. The tidy cubes of brain tissue and the connectivity are represented as nodes and edges respectively. PCA is applied to the functional magnetic resonance imaging data to generate note attributes. If two tidy cubes show similar degree of activation, they will be connected by an edge.
	
	\item {\emph{DBLP}} is a citation network, consisting of bibliography data from computer science. In our experiment, only the authors with at least three publications between 1995 to 2010 are collected. Each author is viewed as a node, and the corresponding titles and abstracts are processed to be the attribute of nodes. Specifically, all titles and abstracts published by an author are concatenated in reverse chronological order. We then pass the concatenated words into the pre-trained BERT$_{BASE}$ \cite{devlin2019bert} and use the vectors of [CLS] in the last layer as the representations. Since the max length of input token for BERT$_{BASE}$ is 512, words that exceed this length limit are removed. The ground-truth category that an author belongs to is decided by the avenues where most of his/her papers are published \footnote{1) Computer Architecture: PPoPP, DAC, MICRO, PODC; 2) Computer Network: SIGCOMM, MobiCom, INFOCOM, SenSys; 3) Data Mining: SIGMOD, ICDE, SIGIR; 4) Computer Theory: STOC, SODA, CAV, FOCS; 5) Multi-Media: SIGGPAPH, IEEEVIS, ICASSP; 6) Artificial Intelligence: IJCAI, ACL, NeurlPS; 7) Computer-Human Interaction: IUI, PERCOM, HCI}.
	
	\item {\emph{ACM}} is similar to the DBLP dataset. Here, only the authors who published at least three papers over the years between 1991 to 2009 are taken into account. Similarly, BERT$_{BASE}$ \cite{devlin2019bert} is applied to generate attributes for each node.
	
\end{itemize}
\begin{table}[!tp]
	\small
	\centering
		\begin{tabular}{c|cccccc}
		\toprule
		Datasets  & {$\#(\text{Vertices})$} & {$*(\text{Vertices})$} & {$\#(\text{Edges})$} & {$*(\text{Edges})$} & {Timestamp} & {Categories} \\
		\midrule
		MOOC & 1382 & 7047 & 12560 & 104642 & 20 & -\\
		Brain & 5000 & 5000 & 74547 & 878207 & 12 & -\\
		DBLP & 20252 & 189296 & 27263 & 1079777 & 16 & 7 \\
		ACM & 10971 & 146040 & 11376 & 636527 & 19 & -\\
		\bottomrule
	\end{tabular}

	\caption{Statistics of datasets. $\#(\cdot)$ and $*(\cdot)$ denote the number at the first and last timestamp.}
	\label{table:datasets}
\end{table}

\paragraph{Baselines} For comparisons, several baseline methods are considered, including both static and dynamic methods.
\begin{itemize}
	\item {\emph{Static Methods:}} DeepWalk \cite{perozzi2014deepwalk}, LINE \cite{tang2015line}, Node2Vec \cite{grover2016node2vec}, CANE \cite{tu2017cane}, SAG \cite{hamilton2017inductive}, WANE \cite{shen2018improved}
	\item {\emph{Dynamic Methods:}} DynTriad\cite{zhou2018dynamic}, DynGEM \cite{goyal2018dyngem}, STAR \cite{xu2019spatio}, tNodeEmbed \cite{singer2019node}, DynAERNN \cite{goyal2020dyngraph2vec}.
	
\end{itemize}
Among the static baselines, the DeepWalk, LINE, and Node2Vec only use the network structure, while the CANE, WANE, and SAGE leverage both the network structure and attributes. When the static methods are used for link prediction of dynamic networks, we apply the static methods to the  network observed until most recently, and the obtained embeddings are then used to predict the links at the next timestamp.

\begin{table}[!ht]
	\centering
	\small
	\setlength{\tabcolsep}{3mm}{
		\begin{tabular}{lccccccccc}
			\toprule
			\multirow{2}{*}{Method} & \multicolumn{3}{c}{DBLP} & \multicolumn{3}{c}{ACM} \\  & ROC-AUC & F1 & PR-AUC & ROC-AUC & F1 & PR-AUC\\
			\midrule
			DeepWalk & 78.26 & 70.14 & 60.13 & 73.53 & 66.12&  56.73  \\ 
			LINE & 74.66 & 67.99 & 58.81 & 70.40 & 64.48 & 55.99  \\ 
			Node2Vec & 77.77 & 69.71 & 60.21 & 72.86 & 65.67 & 56.48  \\ 
			\midrule
			CANE & 79.08 & 72.00 & 61.84 & 77.21 & 70.25 & 60.22 \\
			WANE & 79.55 & 72.26 & 62.05 & 77.40 & 70.37 & 60.35 \\ 
			SAGE & 82.49 & 74.60 & 65.08  & 80.15 & 72.81 & 62.01 \\ 
			\midrule
			DynTriad & 77.41 & 71.36 & 60.32 & 75.59 & 69.60 & 59.45 \\
			DynGEM & 81.54 & 75.65 & 67.27 & 80.04 & 73.85 & 65.69 \\
			STAR & 82.52 & 76.24 & 68.94 & 80.74 & 76.58 & 68.84 \\
			tNodeEmbed & 83.42 & 46.58 & 69.27 & 81.65 & 77.12 & 68.91\\
			DynAERNN & 84.05 & 78.98 & 70.02 & 82.37 & 77.58 & 69.03 \\
			\midrule
			Dane-RNN & 87.14 & 81.53 & 71.42 & {\textbf{86.69}} & 79.51 & 69.87 \\
			Dane-LSTM & 87.55 & {\textbf{82.03}} & 71.78 & 85.64 & 78.71 & 69.12 \\
			Dane-ATT & {\textbf{87.70}} &  81.24 &
			{\textbf{72.27}} & 86.32 & {\textbf{79.89}} &
			{\textbf{70.06}} \\
			\bottomrule
	\end{tabular}}
	\caption{Performance of dynamic link prediction in percentages on DBLP and ACM datasets.}
	\label{table:dlp1}
\end{table}

\begin{table}[!ht]
	\centering
	\small
		\setlength{\tabcolsep}{3mm}{
		\begin{tabular}{lccccccccc}
			\toprule
			\multirow{2}{*}{Method} & \multicolumn{3}{c}{MOOC} & \multicolumn{3}{c}{Brain} \\  & ROC-AUC & F1 & PR-AUC & ROC-AUC & F1 & PR-AUC\\
			\midrule
			DeepWalk & 58.43 & 56.39 & 53.41 & 55.00 & 53.84 & 52.11\\ 
			LINE & 59.02 & 56.95 & 54.76 & 60.14 & 57.96 & 52.89\\ 
			Node2Vec & 57.75 & 55.93 & 53.69 & 57.45 & 55.71 & 53.37 \\ 
			\midrule
			CANE & 61.78 & 58.92 & 55.14 & 61.52 & 57.81 & 54.05 \\
			WANE & 60.93 & 58.47 & 54.95 & 60.78 & 57.74 & 53.61\\ 
			SAGE & 62.59 & 58.64 & 56.00 & 62.07 & 58.54 & 55.13 \\ 
			\midrule
			DynTriad & 57.14 & 54.81 & 52.86 & 54.05 & 52.95 & 49.76 \\
			DynGEM & 62.65 & 57.68 & 55.01 & 62.84 & 57.82 & 54.98\\
			STAR & 63.81 & 59.16 & 55.98 & 63.47 & 58.94 & 56.02\\
			tNodeEmbed & 61.74 & 58.76 & 55.47 & 61.57 & 57.62 & 55.16\\
			DynAERNN & 63.45 & 59.04 & 56.06 & 63.47 & 58.27 & 56.34\\
			\midrule
			Dane-RNN & 64.58 & 58.91 & 57.12 & 64.13 & 58.98 & 56.87\\
			Dane-LSTM & 64.14 & 58.73 & 56.98 & 64.32 & 59.02 & 56.92\\
			Dane-ATT & {\textbf{65.02}} & {\textbf{60.24}} & {\textbf{57.29}} & {\textbf{64.59}} & {\textbf{59.14}} & {\textbf{57.03}}\\
			\bottomrule
	\end{tabular}}
	\caption{Performance of dynamic link prediction in percentages on MOOC and Brain datasets.}
\label{table:dlp2}
\end{table}

\paragraph{Evaluation Metrics} For the task of dynamic link prediction, the widely used evaluation metrics, the area under the ROC curve  (ROC-AUC) \cite{hanley1982meaning}, PR curve  (PR-AUC) \cite{davis2006relationship} and F1 scores,  are utilized to evaluate the performance of learned embeddings. For the dynamic node classification task, a logistic regression model is used to classify the embeddings into different categories. The classifier is trained with the provided labels of nodes. The weighted sum of F1 scores from different categories is used as the performance criteria of this task. All the experiments in this paper are repeated 10 times, and the average results are reported.

\subsection{Dynamic Link Prediction}
For the performance evaluation of link prediction, as done in \cite{goyal2018dyngem}, 20\% of the new links at timestamp $t+1$ are randomly selected  to fine-tune the proposed model, and the rest 80\% are held out for testing. To be fair, the selected 20\% links are also included in training dataset for all the baseline methods. 
The ROC-AUC, PR-AUC, and F1 scores of different models on DBLP, ACM, MOOC and Brain  datasets are shown in Table \ref{table:dlp1} and Table \ref{table:dlp2}, respectively, with the best performance highlighted in bold. Note that the Dane-RNN, Dane-LSTM, and Dane-ATT represent the proposed network embedding models that employ RNN, LSTM, and attentions in the next-timestamp prediction, respectively.
\begin{figure}[ht]
	\centering
	\begin{minipage}[t]{0.48\linewidth}
		\centering
		\includegraphics[scale=0.4]{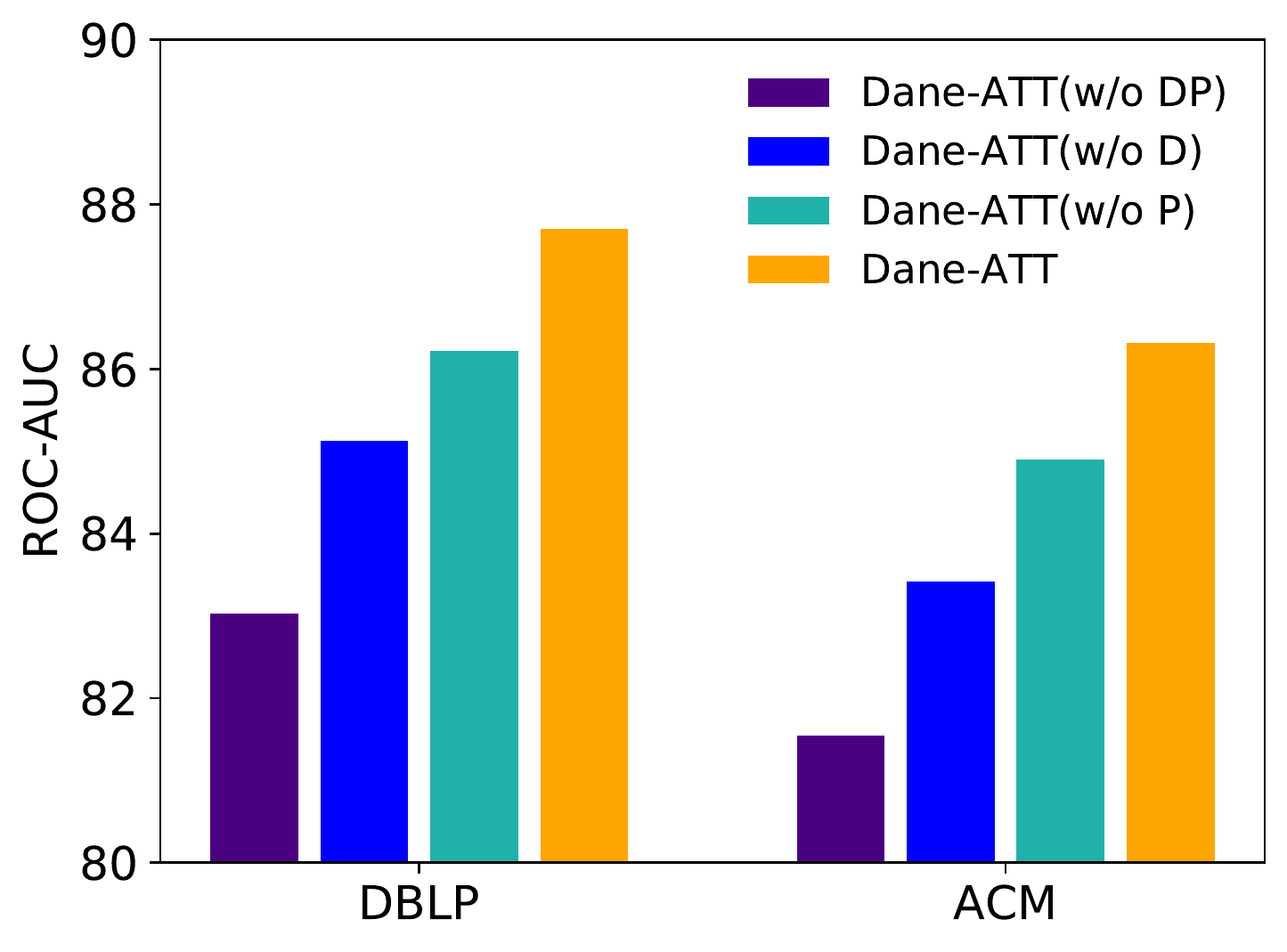}
	\end{minipage}
	\quad
	\begin{minipage}[t]{0.48\linewidth}
		\centering
		\includegraphics[scale=0.4]{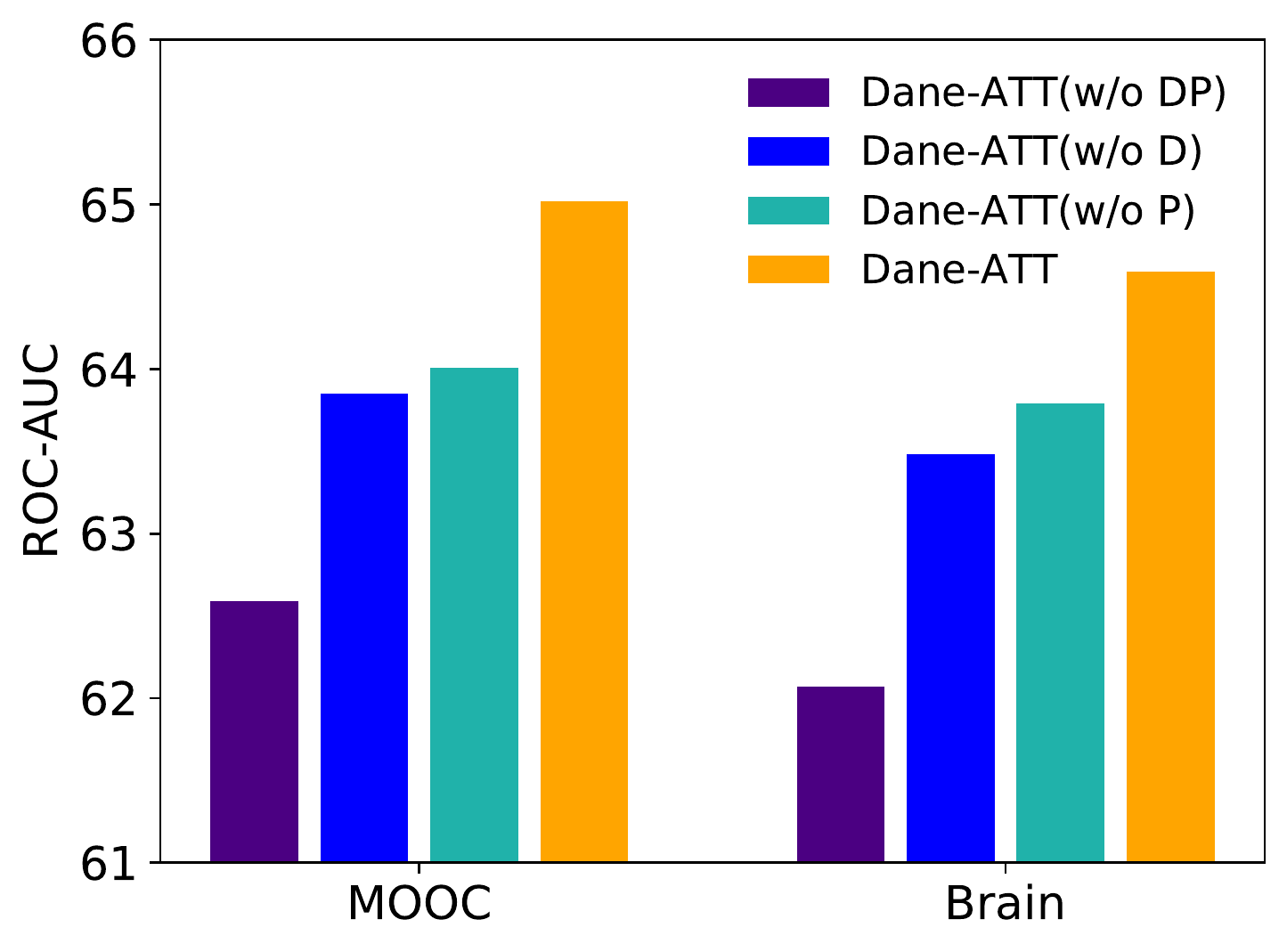}
	\end{minipage}
	\caption{Performance of the variants of Dane-ATT that exclude the activeness vector, dynamic modeling or both.}
	\label{fig:ablation}
\vspace{-2mm}
\end{figure}

From Table \ref{table:dlp1}, it can be seen that the proposed Dane models consistently outperform the baseline
%
methods by a substantial margin on the two considered DBLP and ACM datasets on citation networks. The results suggest that our proposed methods successfully incorporate the network evolution into the embeddings and thus significantly improve the performance on the task of dynamic link prediction. We can also see that the simple attention-based model Dane-ATT achieves a comparable or even better performance than the more complicated Dane-RNN and Dane-LSTM models. This confirms our hypothesis that for the dynamic networks which do not evolve too fast, it is sufficient to employ the attention mechanism to model the temporal dynamics. By examing the static methods, it can be seen that the methods using attributes ({\it e.g.} CANE and SAGE) generally perform better than those that do not ({\it e.g.} DeepWalk), indicating that it is rewarding to incorporate the attributes into the embeddings. We can also observe that
\begin{wrapfigure}{r}{8cm}
	\centering
	\includegraphics[width=1.5in]{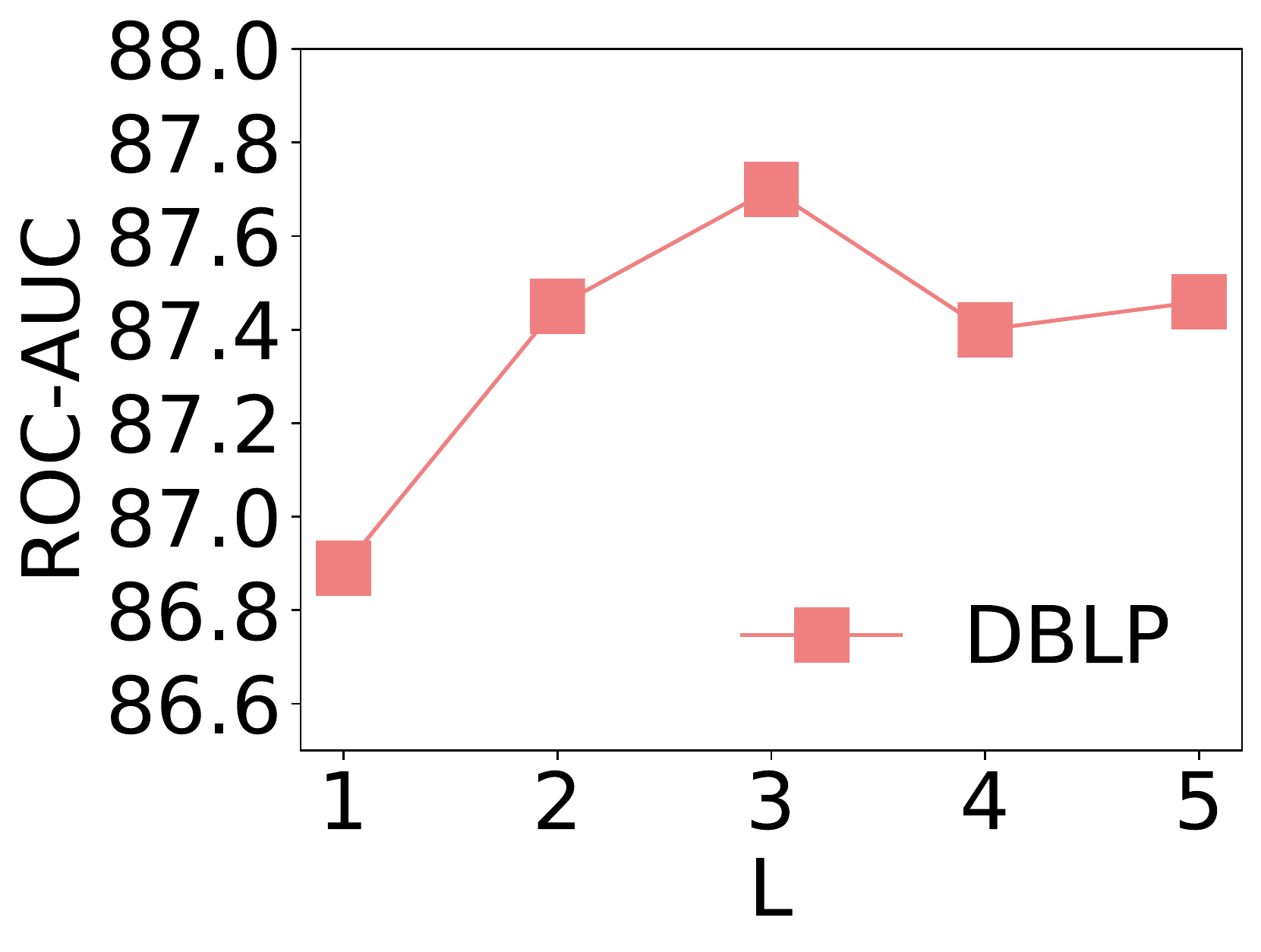}
	\includegraphics[width=1.55in]{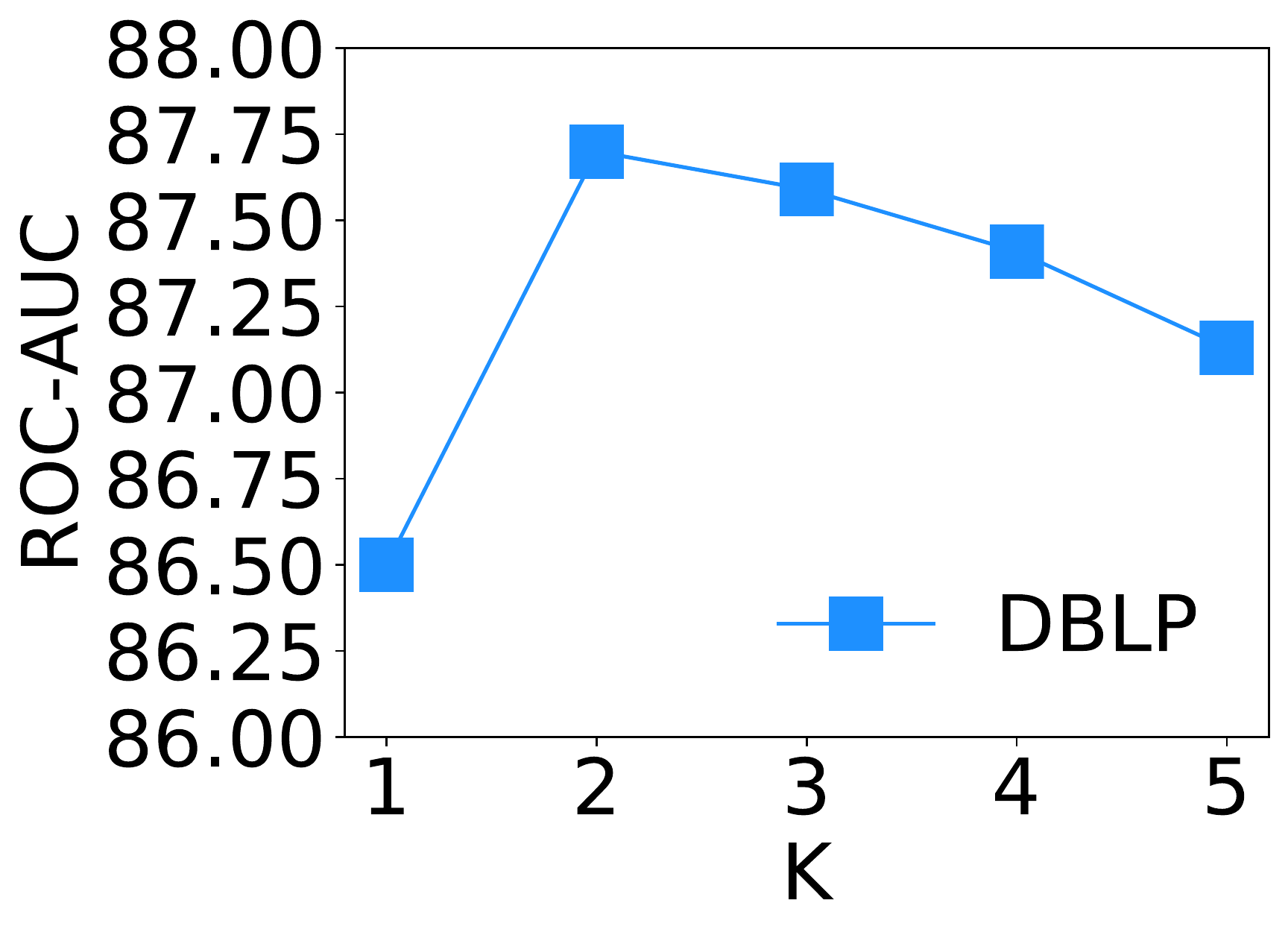}
	
	\includegraphics[width=1.48in]{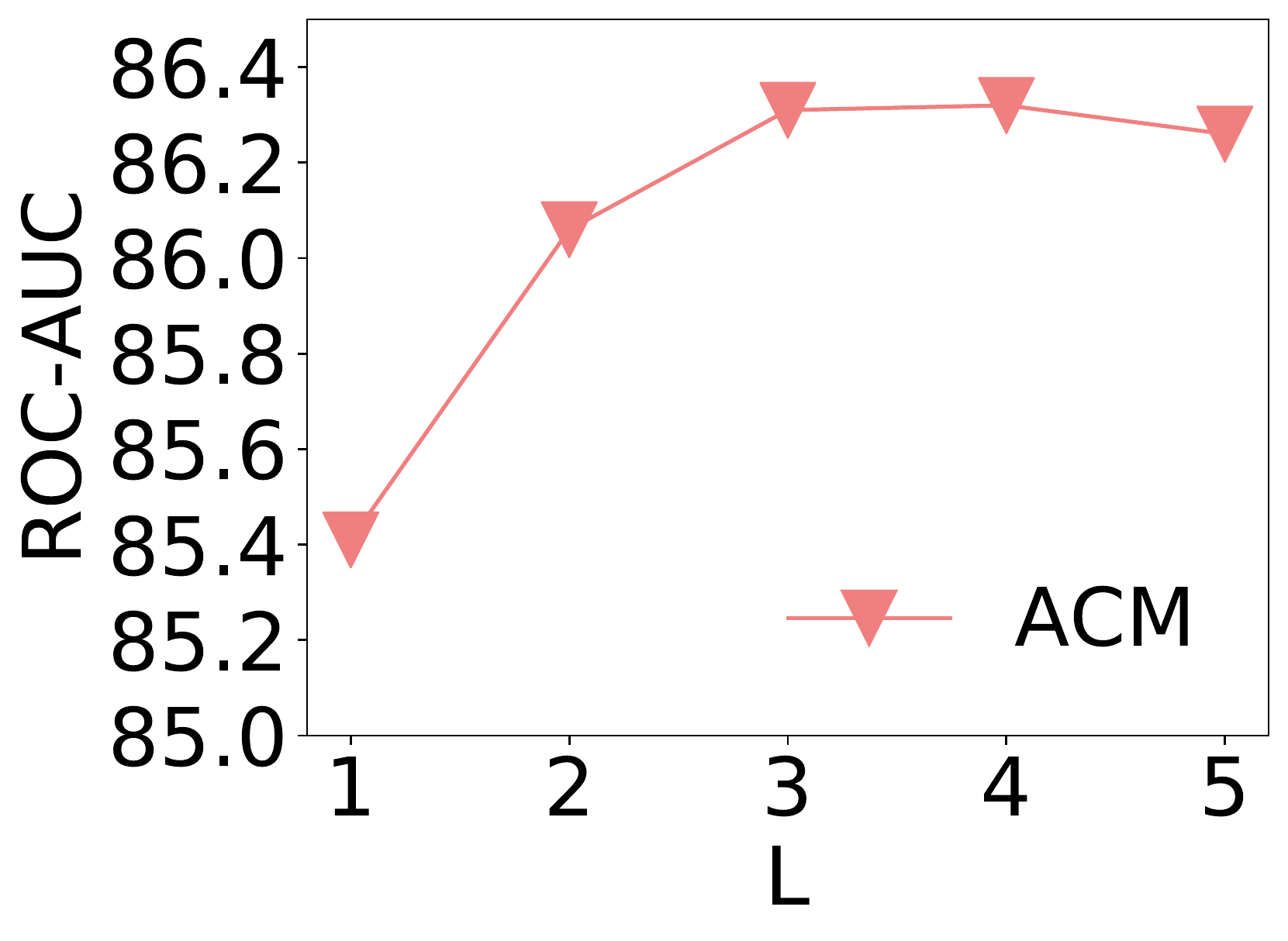}
	\includegraphics[width=1.53in]{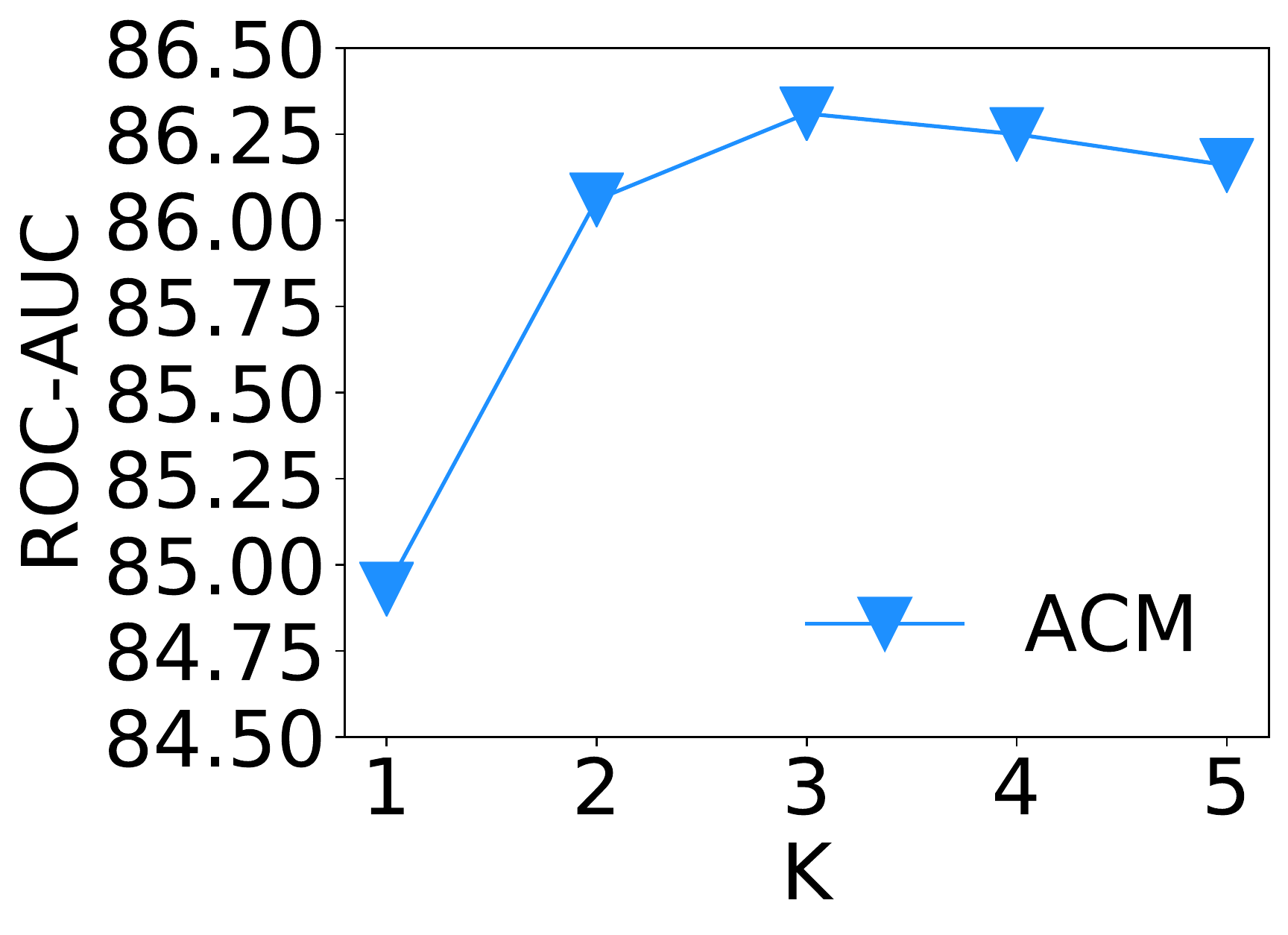}
	\caption{ROC-AUC of Dane-ATT under different values of $L$ and $K$ on the DBLP and ACM datasets.}
	\label{fig:parameterSensitivity}
\end{wrapfigure}
the existing dynamic embedding methods DynGEM and DynAERNN achieve better performance than static methods, although the attributes are not exploited in these dynamic methods, demonstrating the importance of modeling the temporal correlations for future evolvement prediction. By jointly considering the attributes and dynamics, we can see that the proposed Dane models perform best. To further test the model's generalization ability on datasets from other domains, experiments on user action network and brain activity network are conducted. It can be observed from Table \ref{table:dlp2} that the proposed Dane models perform best on MOOC and Brain datasets, which confirms the outstanding generalization capacity of our proposed model.


\paragraph{Impacts of Different Modules} To investigate the importance the activeness-aware mechanism module and temporal correlation modeling module, we evaluate the performance of models that exclude one or both of them, respectively. Specifically, in addition to the Dane-ATT, we consider another three variants: 1) Dane-ATT (w/o P), the model without using the activeness vector; 2) Dane-ATT (w/o D), the model without modeling the dynamics; 3) Dane-ATT (w/o DP), the model that does not use both. The ROC-AUCs evaluated on the four datasets are reported in Fig.\ref{fig:ablation}. It can be seen that without using the activeness vector and the dynamic modeling, an immediate performance drop is observed. This demonstrates the importance of considering both the node activeness spatially and the time correlations temporally. We can see that the drop caused by excluding the dynamic modeling is more significant, suggesting the importance of taking the historical information into account when embedding the dynamic networks. Moreover, if both the activeness vector and dynamic modeling are not used, the worst performance is observed. It is interesting to point out that the Dane-ATT (w/o DP) is actually the static GraphSAGE method, while the Dane-ATT (w/o D) is the static embedding method that has used the activeness of nodes. By comparing the performance of the two variants, the benefits of considering the activeness of nodes are confirmed again.

\paragraph{Impacts of the Parameters $L$ and $K$} The parameter $L$ represents the number of layers used in the neighborhood embedding, while $K$ means how many timestamps that we will look back for the next-time prediction. To investigate the impacts of the two parameters, performances of Dane-ATT with different number of layers $L$ and lookback timestamps $K$ on DBLP and ACM dataset are evaluated. The values of ROC-AUC as functions of $L$ and $K$ are illustrated in Fig.\ref{fig:parameterSensitivity}. It can be seen that as $L$ increases, the performance of proposed model increases rapidly at the beginning and then converges at around $L=3$. The significant improvement at the beginning suggests that incorporating information from high-order neighborhood into the embedding is highly beneficial to the modeling of dynamic evolvement. But as $L$ continues to increase, the improvement is lost. This may be because larger $L$ also results in the decrease of local neighborhood information. Similar trend can be observed in experiments with different $k$. That is, the performance increases as $k$ getting large initially, but then soon gets saturated. This reveals that in the dynamic citation network, it is sufficient to only look back several timestamps when predicting the future trends. This also explains why the attention-based model generally performs better than the RNNs- or LSTM-based ones in the prediction of dynamic networks. That is because the dynamic networks often have a short memory and thereby simple temporal correlations. Thus, for applications of this kind, the more complicated RNN or LSTM may have a detrimental impact on the prediction performance. We believe that when more complex dynamic networks are considered, larger $K$ should be used.

%
%

\begin{figure}[!t]
	\centering
		\begin{minipage}[t]{0.45\linewidth}
			\centering
			\includegraphics[scale=0.4]{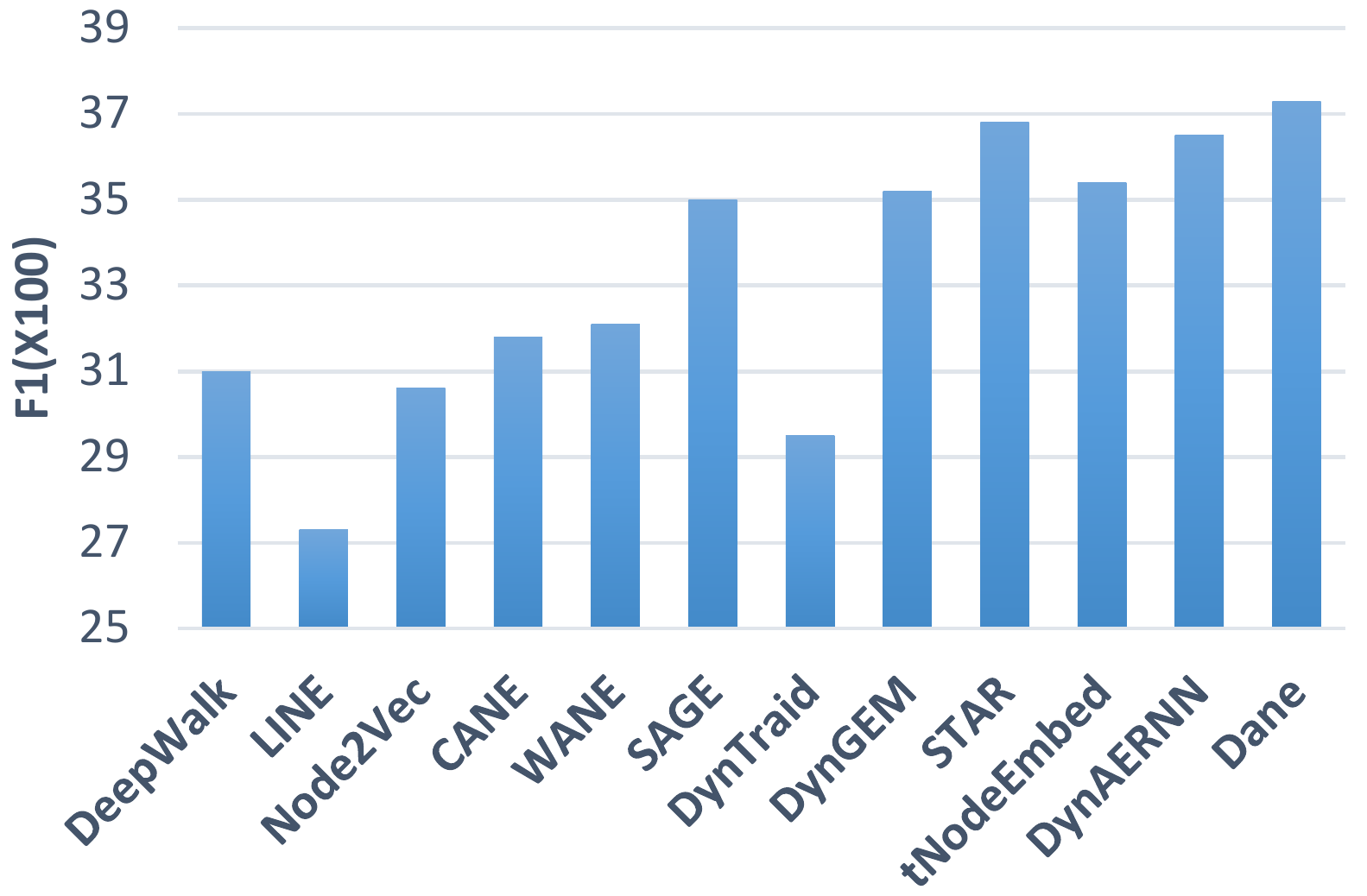}
			\caption{Node classification result on the DBLP dataset}
			\label{fig:bar}
		\end{minipage}
	\quad
		\begin{minipage}[t]{0.45\linewidth}
			\centering
			\includegraphics[scale=0.26]{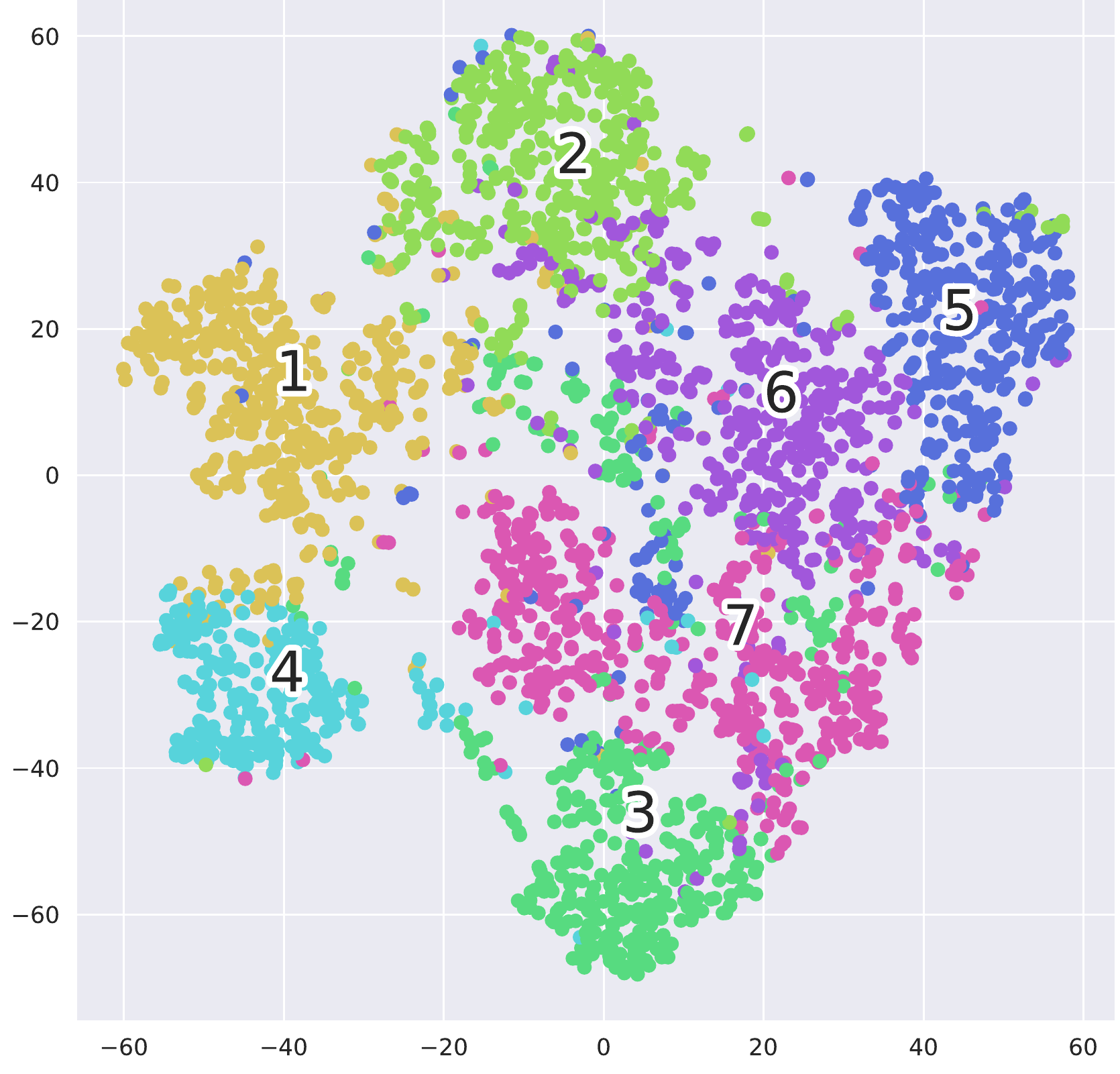}
			\caption{$t$-SNE visualization of our learned network embeddings on DBLP dataset.}
			\label{fig:tsne}
		\end{minipage}%
\vspace{-4mm}
\end{figure}

\subsection{Dynamic Node Classification}
In this section, the experiment of dynamic node classification on the DBLP dataset is conducted. The intuition is that if a model is able to capture the evolution of dynamic attributed networks, the evolution of nodes' categories should also be manifested in the predictive embeddings. To this end, nodes are split into a training and testing set randomly with a proportion of 50\%-50\%. Then a logistic regression classifier regularized by $L_2$ is trained on the node embeddings and their corresponding categories. By noticing that the categories are evolving over time in the DBLP dataset, only users whose categories are changed at the next timestamp are used for testing. 
We repeated the experiment 10 times and the average of weighted sum of F1 scores on different categories are reported. It can be seen from Fig.\ref{fig:bar} that Dane-ATT performs the best among all compared methods considered.  To evaluate the quality of the obtained embeddings, we further visualize them on a 2-D plane with the $t$-SNE. Following \cite{zuo2018embedding}, a sample of 500 nodes for each category is randomly selected and the result is shown in Fig.\ref{fig:tsne}. As shown in Fig.\ref{fig:tsne}, nodes from different categories are separated pretty well, demonstrating that the obtained embeddings preserve the category information of the network well. 
\vspace{-1mm}

\section{Conclusion}
In this paper, a dynamic attribute network embedding framework is proposed to track the network evolution by modeling the high-order correlations in spatial and temporal dimensions jointly. To this end, an activeness-aware neighborhood embedding method is proposed to extract the high-order neighborhood information at each timestamp. Then, an embedding prediction framework is developed to capture the temporal correlations. Extensive experiments were conducted on four real-world datasets over the tasks of link prediction and node classification, confirming the ability of the model to track the evolutions of dynamic networks.

\section*{Acknowledgements}
This work is supported by the National Natural Science Foundation of China (No. 61806223, 61906217, U1811264), Key R\&D Program of Guangdong Province (No. 2018B010107005) and Fundamental Research Funds for the Central Universities (No. 191gjc04).


\clearpage
\bibliographystyle{coling}
\bibliography{coling2020}

\end{document}